\documentclass[amsmath,amssymb,twocolumn]{revtex4}
\usepackage{graphicx}
\usepackage{dcolumn}
\usepackage{bm}
\usepackage{amsmath,mathrsfs}
\usepackage{graphicx,epsfig}
\usepackage{subfigure}
\usepackage{hyperref}
\usepackage{epsf}
\usepackage{color}

\begin{document}
%
%
\title{{\bf Spatially varying modified Chaplygin gas wormhole with minimum violation of energy condition}}
\author{S. Rastgoo}\email{rastgoo@sirjantech.ac.ir}
\author{ F. Parsaei }\email{fparsaei@gmail.com}
%
%
\affiliation{ Physics Department , Sirjan University of Technology, Sirjan 78137, Iran.}
%
%




\begin{abstract}
\par
In this study, we explore the existence of traversable wormhole solutions in general relativity using a power-law shape function given by $b(r) = r^{\alpha}$. Departing from the conventional approach—where matter content is derived a posteriori from the field equations—we instead prescribe a generalized, radially dependent equation of state: $p(r) = \omega_1(r)\rho(r) - \frac{\omega_2(r)}{\rho(r)^{\gamma}}$. This formulation smoothly transitions from a Chaplygin-like behavior at small radii to a linear barotropic regime at large distances. We obtain an analytical expression for the energy density and derive the radial pressure in closed form. The redshift function is derived using the field equation and the equation of state. Ultimately, we demonstrate that this spatially varying modified Chaplygin gas equation of state successfully confines the exotic matter to the immediate vicinity of the throat. Moreover, we demonstrate that the requirement of consistency between the prescribed equation of state and the Einstein field equations fixes the throat value of the second coefficient, $B =\alpha$, thereby removing the apparent  indeterminacy in the redshift function near the throat. For the specific choices $\omega_1(r) = \left(1 - r^{-m}\right)\omega_\infty$ and $\omega_2(r) = B\, r^{-k}$, with the decay condition $k > \gamma(3-\alpha)$, the non-linear term is suppressed at spatial infinity and the model reduces to a barotropic fluid with $p \simeq \omega_\infty \rho$, yielding an asymptotically well-behaved geometry. Our results indicate that the choice of the equation of state is not a mere mathematical convenience but a fundamental ingredient that controls both the localization of exotic matter and the asymptotic viability of the wormhole solution.\\
\end{abstract}

\maketitle
\section{Introduction}

Representing hypothetical bridges between distant regions of spacetime or entirely separate universes, traversable wormholes stand as some of the most conceptually profound solutions derived from Einstein’s field equations \cite{WH}. The study of traversable wormholes is deeply intertwined with core inquiries into causality, the topology of spacetime, and the theoretical viability of topological shortcuts. Determining what kind of matter can support traversable wormholes is a central challenge in the field, as these structures typically demand a violation of the null energy condition (NEC) \cite{Visser}. The accelerated expansion of the universe is one of contemporary cosmology's most intriguing discoveries \cite{expan}. This acceleration is driven by a dominant dark energy component characterized by the equation of state (EoS) $p = \omega \rho$ ($\omega < -1/3$). Although $\omega < -1$ (the phantom energy state) seems exotic in a lab setting, it is fully supported by cosmological data \cite{apli}. Phantom energy may theoretically make traversable wormholes possible. Phantom wormhole solutions have been investigated in the literature, specifically focusing on thin-shell \cite{phantom0, phantom, phantom2} and asymptotically flat \cite{phantom01, phantom11, phantom1, phantomH, phantom4, LPR} models. In the context of modified gravity, wormholes governed by a linear EoS have also been explored as a means to minimize—or entirely circumvent—the requirement for exotic matter \cite{b1, b2, b3, Cartan1,  Rast, Rast1, fq, fq2, fq4, fq44, fq5,fR11, fR22, fR55, foad3, SR2, SR1, foad4, fq-sara, Sodab, Sara}.

The phantom EoS diverges fundamentally from Chaplygin gas (CG) models in its underlying physics and cosmological implications. By violating the NEC, phantom models risk a catastrophic Big Rip singularity in the future and frequently encounter ghost instabilities in field-theoretic frameworks \cite{Chap0}. Conversely, the generalized Chaplygin gas (GCG), given by $p = -B/\rho^\gamma$, seamlessly bridges the matter-dominated era and late-time accelerated expansion, ultimately behaving like a cosmological constant while generally avoiding the phantom regime \cite{Chap0}. The modified Chaplygin gas (MCG)—expressed as $p = A\rho - B/\rho^\gamma$—extends this versatility by using its linear term to mimic radiation or matter, while the nonlinear, negative-pressure component drives acceleration. Consequently, Chaplygin-type models offer a more adaptable framework for tracking cosmic evolution than a strictly phantom fluid, though they are not without challenges. They must contend with stringent observational constraints, as well as potential issues involving sound speed, density perturbations, and structure formation. At the same time, specific parameter tunings can allow these models to mimic phantom behavior or bypass the Big Rip entirely. Ultimately, the primary strength of the Chaplygin and modified Chaplygin EoS lies in their capacity to unify multiple epochs of cosmic history into a single fluid description, provided that individual model stabilities and observational viabilities are rigorously verified.

Lobo investigated exact solutions for spherically symmetric, traversable wormholes supported by the GCG, suggesting they may arise from density fluctuations in the GCG cosmological background \cite{Chap0}. The accretion of a GCG onto wormholes is investigated in \cite{Chap08}.
 Spherical thin-shell wormholes constructed with a pure CG were first explored by Eiroa and Simeone \cite{Chap1}. Building on this, Bandyopadhyay et al. \cite{Chap2} demonstrated that stable static wormhole solutions can be achieved in both Schwarzschild and Schwarzschild-de Sitter backgrounds using a MCG. Furthermore, investigations of the Tolman–Oppenheimer–Volkoff equations by other researchers \cite{Chap3, Chap4} revealed wormhole-like solutions containing spacetime singularities when frameworks are filled with CG and GCG \cite{jamil2}. Eiroa \cite{Chap5} later examined GCG thin-shell wormholes, finding that their static solutions are largely unstable. Additionally, utilizing the cut-and-paste method, Sharif and Azam \cite{Chap6} constructed spherical thin-shell wormholes and evaluated their stability near a MCG. Amirabi et al. \cite{Chap07} investigated the effect of the Gauss-Bonnet parameter on the stability of thin-shell wormholes with a GCG equation of state. In \cite{Chap10}, spherically symmetric thin-shell wormholes—supported by a GCG within Einstein gravity coupled to Born-Infeld electrodynamics—are constructed, and their stability under radial perturbations is analyzed.
 Jamill et al. investigated wormholes supported by polytropic phantom energy, a generalized form of phantom energy that occasionally encompasses CG models  \cite{mobasher}.
The study of noncommutative geometry-inspired wormholes supported by quasi-de Sitter and Chaplygin-like EoS was  presented in \cite{Chap09}. Other researchers have investigated the wormhole utilizing the CG or MCG equation of state \cite{Chap7, Chap8, Chap11}

Considerable attention has been dedicated to minimizing exotic matter to ensure the physical viability of wormholes. In this context, the EoS chosen for the matter at the wormhole throat is critical for maintaining static solutions. To address the challenge of asymptotic flatness or to limit the required amount of exotic matter, some researchers employ the "cut-and-paste" method, wherein the interior wormhole solution is matched with an exterior Schwarzschild metric \cite{cut, cut1, cut2, cut3}. Although widely used to localize exotic matter and achieve asymptotic flatness, this approach is often considered physically unrealistic.
In the framework of GR, variable EoS models \cite{Remo, variable, variable2}, polynomial EoS formulations \cite{foad}, and polytropic wormholes \cite{ P1, P2, P22, P3, P0, P4} employ a variable EoS to reduce exotic matter violations. Building on past findings, this paper explores the geometry and configuration of exotic matter in GR by introducing a spatially varying modified Chaplygin gas(SV-MCG) EoS.

The remainder of this paper is organized as follows. Section \ref{sec2} outlines the wormhole geometry and derives the field equations. Section \ref{sec3} introduces the EoS and formulates general wormhole solutions for inhomogeneous media. Concluding remarks are provided in Section \ref{sec4}. Throughout this work, geometrized natural units are used, setting $8\pi G = c = 1$.

\section{ Wormhole framework} \label{sec2}
In this section, we formulate a general framework for the construction of static spherical wormholes. The line element for a spherically symmetric, static wormhole is given by
\begin{equation}\label{1}
ds^2=-\exp(2\Phi(r))dt^2+\frac{dr^2}{1-\frac{b(r)}{r}}+r^2(d\theta^2+\sin^2\theta
d\phi^2),
\end{equation}
where $\Phi(r)$ represents the gravitational redshift function, and $b(r)$ is the shape function that dictates the spatial geometry of the wormhole. The spacetime geometry must fulfill the flaring-out criterion at or adjacent to the throat, denoted as
\begin{equation}\label{2}
b'(r_0)<1.
\end{equation}
This condition is essential for the wormhole to flare outward and bridge two distinct asymptotic regions. At the wormhole throat, located at $r = r_0$, the function satisfies
\begin{equation}\label{3}
b(r_0)=r_0.
\end{equation}
To remain traversable, wormholes must be devoid of event horizons. Specifically, the shape function must satisfy
\begin{equation}\label{4}
b(r)<r,\ \ {\rm for} \ \ r>r_0
\end{equation}
and the redshift function, $\Phi(r)$, must stay finite throughout spacetime. Furthermore, consistency with large-scale cosmological observations requires that any physically viable wormhole be asymptotically flat, meaning:
\begin{equation}\label{5}
\lim_{r\rightarrow \infty}\frac{b(r)}{r}=0,\qquad   \lim_{r\rightarrow \infty}\Phi(r)=0.
\end{equation}

We naturally model the matter content as an anisotropic fluid with the stress-energy tensor $T^{\mu}_{\phantom{\mu}\nu} = \text{diag}[-\rho(r), p_r(r), p_t(r), p_t(r)]$, defined by the energy density $\rho(r)$, radial pressure $p_r(r)$, and tangential pressure $p_t(r)$. Based on line element (\ref{1}), we use the Einstein field equations to determine the following matter distribution:
\begin{eqnarray}
&&
\!\!\!\!\!\!\!\!\!\!\!\!\!\!\!\!\!\!\!
\rho(r)=\frac{b'(r)}{r^{2}}\,,
\label{6}
\\
&&\!\!\!\!\!\!\!\!\!\!\!\!\!\!\!\!\!\!\!
 p_r(r)=\frac{2}{r}\left(1-\frac{b(r)}{r}\right)\Phi'(r)
-\frac{b(r)}{r^{3}}\,,
\label{7}
\\
&&
\!\!\!\!\!\!\!\!\!\!\!\!\!\!\!\!\!\!\!\!\!
 p_t(r)=-\frac{b'(r)r-b(r)}{2r^{2}}
\left(\Phi'(r)+\frac{1}{r}\right)
\nonumber\\
&& \ \ \left(1-\frac{b(r)}{r}\right)
\left[\Phi''(r)+\Phi'(r)\left(\Phi'(r)+\frac{1}{r}\right)\right]\,.
\label{8}
\end{eqnarray}
Throughout this work, primes denote derivatives with respect to $r$. 

The Chaplygin gas EoS, expressed as
 \begin{equation}\label{9}
  p=-\frac{B}{\rho},
   \end{equation}
 where $B>0$ is a positive constant, has garnered significant interest in modern cosmology. This framework offers a unified description of cosmic evolution, seamlessly bridging a matter-dominated era with late-time accelerated expansion. In the limit of high energy densities, the pressure vanishes, causing the fluid to mimic dust-like (pressureless) matter; conversely, at low densities, it asymptotically approaches a negative-pressure state akin to a cosmological constant. Crucially, unlike phantom dark energy models, the standard CG can drive cosmic acceleration while satisfying the NEC. Despite these theoretical advantages, the model encounters notable hurdles regarding cosmological perturbation dynamics, effective sound speed, and large-scale structure formation. These shortcomings have driven the formulation of generalized and modified Chaplygin gas frameworks, which incorporate extra parameters to enhance observational viability and flexibility.

 The modified Chaplygin gas EoS is a generalized form of the standard CG model and is commonly expressed as
\begin{equation}\label{9c}
p(\rho)=A\rho-\frac{B}{\rho^\gamma},
\end{equation}
where  $A$, $B$, and $\gamma$ are model parameters, typically satisfying $B>0$ and $0\leq\gamma\leq1$. The principal difference between the MCG and CG models lies in the additional linear term $A\rho$, which provides greater flexibility in describing the cosmic evolution. In contrast, the standard CG is characterized by the EoS (\ref{9}) which corresponds to the special case $A=0$ and $\gamma=1$ of the MCG model. The CG model can interpolate between a pressureless-matter-like regime at high energy densities and a cosmological-constant-like regime at low densities. The MCG model extends this behavior by allowing the linear term to influence the early-time dynamics, potentially describing different evolutionary phases of the universe, while the inverse-power term can drive late-time accelerated expansion. Consequently, the MCG model offers additional freedom in fitting cosmological observations; however, its physical viability depends on constraints from observational data, perturbation stability, and structure formation. Consequently, the MCG equation of state  functions as a hybrid of the linear and CG equations of state.

In \cite{variable}, we have presented a new class of wormhole solutions by assuming a variable EoS as
\begin{equation}\label{9b}
p_r(r)=\omega(r) \rho(r),
\end{equation}
where
\begin{equation}\label{9d}
\omega(r)=\omega_\infty+g(r),
\end{equation}
and
\begin{equation}\label{9e}
\lim_{r\rightarrow \infty} g(r)=0.
\end{equation}
Following this approach, we introduce a new variable MCG EoS as
\begin{equation}\label{9f}
p_r(r)=\omega_1(r)\rho-\frac{\omega_2(r)}{\rho^\gamma}.
\end{equation}
We assume the conditions
\begin{eqnarray}\label{9g}
\lim_{r\rightarrow r_0} \omega_1(r)=0 \nonumber\\
\lim_{r\rightarrow \infty} \omega_1(r)=\omega_\infty
\end{eqnarray}
and
\begin{eqnarray}\label{9h}
\lim_{r\rightarrow r_0} \omega_2(r)=B \nonumber\\
\lim_{r\rightarrow \infty} \omega_2(r)=0.
\end{eqnarray}

Below, we summarize the properties of the  SV-MCG model:
\begin{itemize}
        \item  Promoting the conventional constant MCG parameters to radial-dependent functions allows independent regulation of matter distribution from the throat to asymptotic regions.
    \item  By imposing $\omega_1(r_0) \to 0$ and $\omega_2(r_0) \to B$, the EoS reduces to $p(r_0) \approx -\frac{B}{\rho(r_0)}$, generating the strong negative pressure required to support the wormhole throat and control NEC violations.
    \item  With $\omega_1(\infty) \to \omega_\infty$ and $\omega_2(\infty) \to 0$, the model smoothly transitions into a standard barotropic fluid $p(\infty) \approx \omega_\infty \rho(\infty)$.
    \item  It successfully bridges the localized exotic requirements of wormhole throats with standard, physically acceptable asymptotic behaviors.
\end{itemize}

However, the system involves five unknown functions of $r$—namely $\rho(r)$, $p_r(r)$, $p_t(r)$, $b(r)$, and $\Phi(r)$. To fully determine the system, one must impose specific constraints on either $b(r)$ or $\Phi(r)$. Alternatively, plausible stress-energy components can be assumed, allowing the metric fields to be derived directly through the field equations. Because gravitational field equations in wormhole physics are underdetermined—meaning the unknown metric and stress-energy components outnumber the independent equations—researchers typically rely on specific closure strategies. These include: \textit{(i)} the zero tidal force (ZTF) or constant redshift ansatz ($\Phi(r) = 0$) to simplify the temporal metric; \textit{(ii)} assuming a physically acceptable shape function $b(r)$ that meets the necessary criteria; \textit{(iii)} applying the inverse reconstruction method, which derives the spacetime geometry from a predefined energy density or matter profile; and \textit{(iv)} using an EoS-driven approach—such as the proposed SV-MCG model—combined with a preset metric function to fully determine the remaining thermodynamic variables and test the ECs.

In this paper, we consider a non-vanishing redshift function alongside the SV-MCG equation of state. This setup yields four equations with five unknown functions. To close the system, we introduce specific, well-known shape functions, enabling us to solve the field equations and determine the redshift function, and ultimately, energy momentum tensor (EMT) components. This methodology is chosen because satisfying the rigorous mathematical conditions required for the wormhole shape function is significantly more demanding than for the remaining functions.
Using  (\ref{6}) and (\ref{7}) leads to
\begin{equation}\label{10}
\Phi(r)=\frac{1}{2}\int\frac{b(r)+r^3\,p_r(r)}{r(r-b(r))}dr
\end{equation}
In the next section, we introduce a viable shape function and use Eq. (\ref{10}) to determine the corresponding redshift functions.

In general relativity (GR), energy conditions (ECs) provide criteria for assessing the physical admissibility of matter distributions through constraints on the stress--energy tensor, \(T_{\mu\nu}\). These conditions play a particularly important role in wormhole physics, where the NEC is closely associated with the geometric requirements for maintaining a traversable throat. For a static wormhole, the flare-out condition generally entails NEC violation at or near the throat within standard GR, implying the presence of matter with properties commonly referred to as exotic. The NEC requires that \(T_{\mu\nu}k^\mu k^\nu\geq 0\) for every null vector \(k^\mu\). The weak energy condition (WEC) further requires that every timelike observer measure a non-negative energy density, expressed as \(T_{\mu\nu}v^\mu v^\nu\geq 0\) for all timelike vectors \(v^\mu\). The dominant energy condition (DEC) supplements these restrictions by requiring the energy density to be non-negative and the energy--momentum flux to be causal. The strong energy condition (SEC), meanwhile, constrains the effective gravitational influence of matter and is commonly associated with the focusing behavior of timelike geodesics through the Raychaudhuri equation. For an anisotropic matter distribution characterized by energy density \(\rho\), radial pressure \(p_r\), and tangential pressure \(p_t\), these conditions can be written as
\begin{align}
\mathrm{NEC}: \quad
&\rho+p_r\geq 0,\qquad
\rho+p_t\geq 0,
\label{EC1}
\\
\mathrm{WEC}: \quad
&\rho\geq 0,\qquad
\rho+p_r\geq 0,\qquad
\rho+p_t\geq 0,
\label{EC2}
\\
\mathrm{DEC}: \quad
&\rho\geq 0,\qquad
\rho-|p_r|\geq 0,\qquad
\rho-|p_t|\geq 0,
\label{EC3}
\\
\mathrm{SEC}: \quad
&\rho+p_r\geq 0,\quad
\rho+p_t\geq 0,\quad
\rho+p_r+2p_t\geq 0.
\label{EC4}
\end{align}
Following the diagnostic approach adopted in Ref.~\cite{fq}, the above inequalities can be examined by introducing the functions
\begin{align}
H(r)&=\rho+p_r,\qquad
H_1(r)=\rho+p_t,\qquad
H_2(r)=\rho-|p_r|,\nonumber\\
H_3(r)&=\rho-|p_t|,\qquad
H_4(r)=\rho+p_r+2p_t.
\label{EC}
\end{align}
The signs of these functions determine whether the corresponding EC inequalities are satisfied at each radial position.

\section{Wormhole solutions }\label{sec3}

Within the framework of traversable wormhole physics, the choice of the shape function $b(r)$ is critical; it determines the gravitational geometry, the distribution of matter at the throat, and the degree to which exotic ECs are violated. In this section, we assume specific shape functions and determine the corresponding redshift function using Eq. (\ref{10}). Subsequently, the EMT components are derived from the field equations, allowing us to analyze the physical properties of the wormhole.

As  a viable choice for $p_r(r)$, we assume
 \begin{eqnarray}
\omega_1(r)=(1-\frac{1}{r^m})\omega_\infty,\label{14a}\\
\omega_2(r)=(\frac{B}{r^k}).\label{14b}
\end{eqnarray}
These functions satisfy conditions (\ref{9g}) and (\ref{9h}). Next, by substituting (\ref{14a}) and (\ref{14b}) into (\ref{10}) along with a known shape function, we determine the redshift function. We assume $\gamma=1$ throughout this work.
Among various analytical approaches, the power-law shape function has proven to be a particularly versatile and robust framework. By parameterizing the function $b(r)$ with power-law exponents, researchers can accurately model the transition from non-trivial throat geometries to asymptotically flat, de Sitter, or anti-de Sitter spacetimes. Additionally, this approach facilitates a systematic analysis of the flaring-out condition ($b'(r) < b(r)/r$), the distribution of the effective stress-energy tensor, and the minimization of exotic matter requirements within various modified gravity theories. Widely regarded as the standard shape function in wormhole physics, it satisfies all fundamental requirements for wormhole construction in both GR and modified gravity. Accordingly, we adopt the following power-law shape function:
\begin{equation}\label{11}
b(r)= r^\alpha, \qquad \alpha<1.
\end{equation}
Substituting the shape function (\ref{11}) in (\ref{6}) yields the density profile
\begin{equation}\label{11a}
\rho(r)= \alpha r^{\alpha-3}.
\end{equation}
Ensuring a positive energy density requires the condition
\begin{equation}\label{11b}
0<\alpha<1.
\end{equation}
Using (\ref{11a}), we obtain
\begin{equation}\label{11bb}
\rho(r_0)= \alpha.
\end{equation}
This implies that $\alpha$ is the energy density at the wormhole throat.
Using (\ref{3}) and (\ref{11}) in (\ref{7}) gives
\begin{equation}\label{11c}
p_r(r_0)= -\frac{1}{r_0^2}=-1.
\end{equation}
Substituting (\ref{11a}) into (\ref{9f}) yields
\begin{equation}\label{11d}
p_r(r_0)= -\frac{B}{\alpha}.
\end{equation}
Comparing (\ref{11c}) and (\ref{11d}) gives
\begin{equation}\label{11e}
\alpha=B.
\end{equation}
This condition is essential for constructing wormhole solutions within the SV-MCG model background. In fact, relation (\ref{11e}) establishes a direct link between the geometric parameters of the system and the behavior of the fluid supporting the wormhole. For a wormhole solution to be physically viable, the radial pressure must vanish at spatial infinity. Applying conditions (\ref{14b}) and (\ref{11a}) therefore yields:
\begin{equation}\label{12a}
k>3-\alpha.
\end{equation}
Assuming $k$ is an integer then substituting (\ref{11}) into (\ref{10}) and applying condition (\ref{11d}) yields
\begin{equation}\label{shape22}
\Phi(r)=\frac{1}{2}(\Phi^{\alpha}_{1}+\omega_{\infty}\alpha \left(\Phi^{\alpha}_{1}-\Phi^{\alpha}_{1+m}\right) -\Phi^{\alpha}_{k-5+2\alpha}).
 \end{equation}
where
\begin{eqnarray}\label{shape23}
\Phi^{\alpha}_{1}&=&\int \frac{1}{r(r^{(1-\alpha)}-1)}dr=\frac{\ln(1-\frac{1}{r^{1-\alpha}})}{1-\alpha},
\nonumber \\
\Phi^{\alpha}_{m}&=&\int \frac{1}{r^m(r^{(1-\alpha)}-1)}dr.
 \end{eqnarray}
  Because the general solutions for $\Phi(r)$ with arbitrary $\alpha$ and $m$ are expressed in terms of hypergeometric functions, we simplify the problem by fixing certain parameters. Specifically, we set $\alpha = \frac{1}{2}$, which satisfies the condition $0 \leq \alpha < 1$ required to ensure a positive energy density. In the subsequent analysis, we examine how the free parameters $m$ and $k$ influence the physical outcomes. Specifically, we set $m = 1$ and $\alpha=\frac{1}{2}$ then the redsift function is
  \begin{equation}\label{12}
\Phi(r) = -\frac{1}{4} \left[ \omega_{\infty} \left( \frac{2}{\sqrt{r}} + \frac{1}{r} \right) + 2 \sum_{i=1}^{k-5} \left( \frac{2}{(2i-1)r^{\frac{2i-1}{2}}} + \frac{1}{i r^i} \right) \right]
\end{equation}
Note that values of $k$ in the range $3-\alpha < k \leq 5$ result in redshift functions that violate asymptotic flatness. Therefore, the applicable range for equation (\ref{12}) is restricted to:
\begin{equation}\label{12c}
k\geq 6.
\end{equation}
With the geometrical framework of the theory now fully established, the EMT components can be determined directly via the field equations.

\begin{figure}
\centering
  \includegraphics[width=3 in]{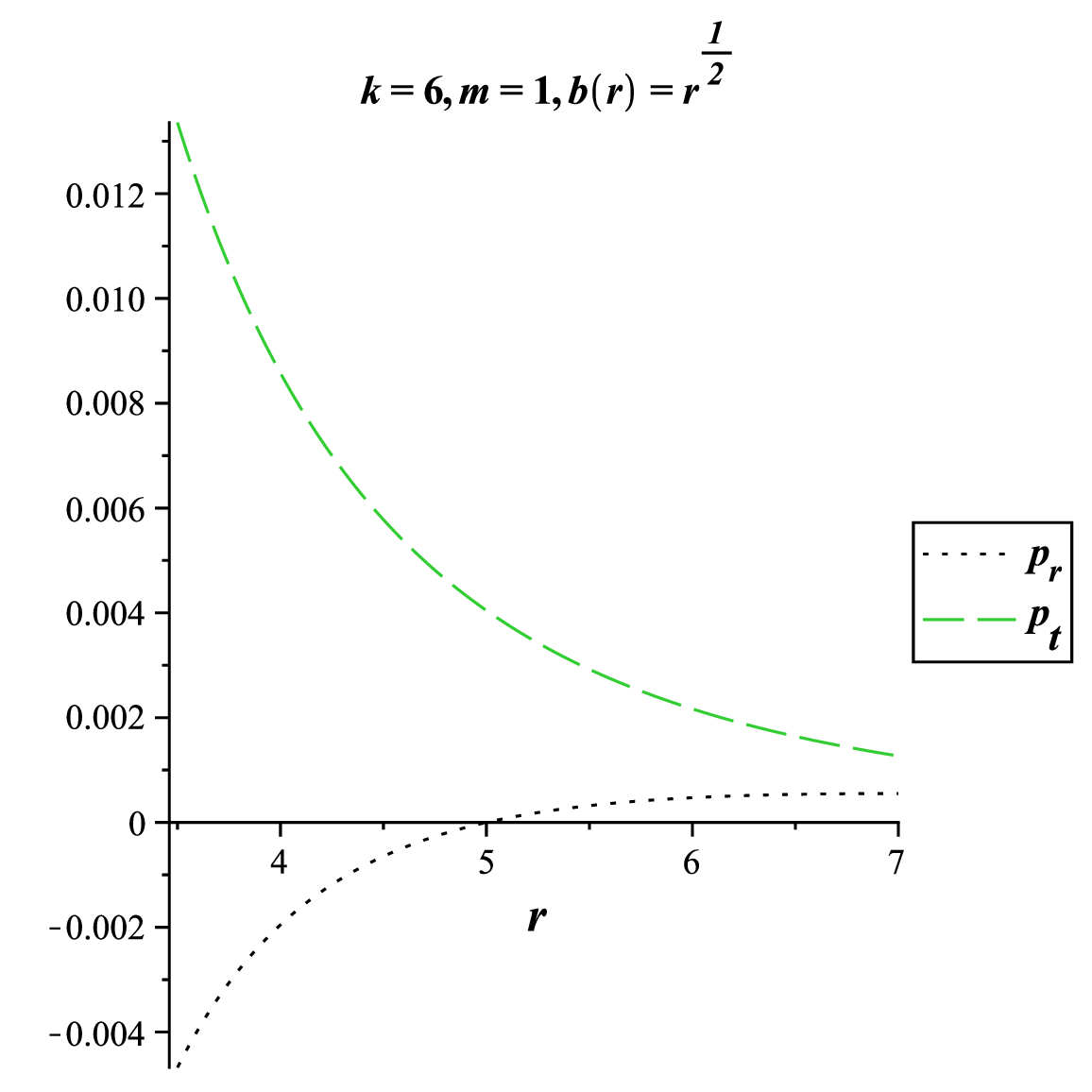}
\caption{Th graph displays the radial pressure $p_r(r)$ (dot curve) and the tangential pressure $p_t(r)$ (dashed curve) as functions of the radial coordinate, for the shape function $b(r)=\sqrt{r}$ with $k=6$ and $m=1$. The tangential pressure remains positive throughout the entire domain, whereas the radial pressure is negative in the vicinity of the throat and changes sign at $r_p=5$. Further details are given in the text.}
 \label{fig1}
\end{figure}

\begin{table*} 
\begin{tabular}{|l|l|l|l|}
\hline
$k$  & $p_{r}(r)=0$ & $H(r)=H_2(r)=0$& $H_3(r)=0$  \\ \hline
 $k=6$  & $r_p=5$ & $r_H=1.6$ & $r_{H_3}=1.63$    \\ \hline
 $k=7$  & $r_p=2.56$ & $r_H=1.3$ &  $r_{H_3}=2.15$   \\ \hline
$k=8$ &  $r_p=2$ & $r_H=1.2$ &   $r_{H_3}=2.14$   \\ \hline
$k=9$ & $r_p=1.74$ &$r_H=1.16$ &  $r_{H_3}=2.07$   \\ \hline
\end{tabular}
\caption{The results for different $\Phi(r)_K$ while $b(r)=\sqrt{r}$, $\omega_\infty=1/2$ and $m=1$.  }\label{Tab1}
\end{table*}

\begin{figure}
\centering
  \includegraphics[width=3 in]{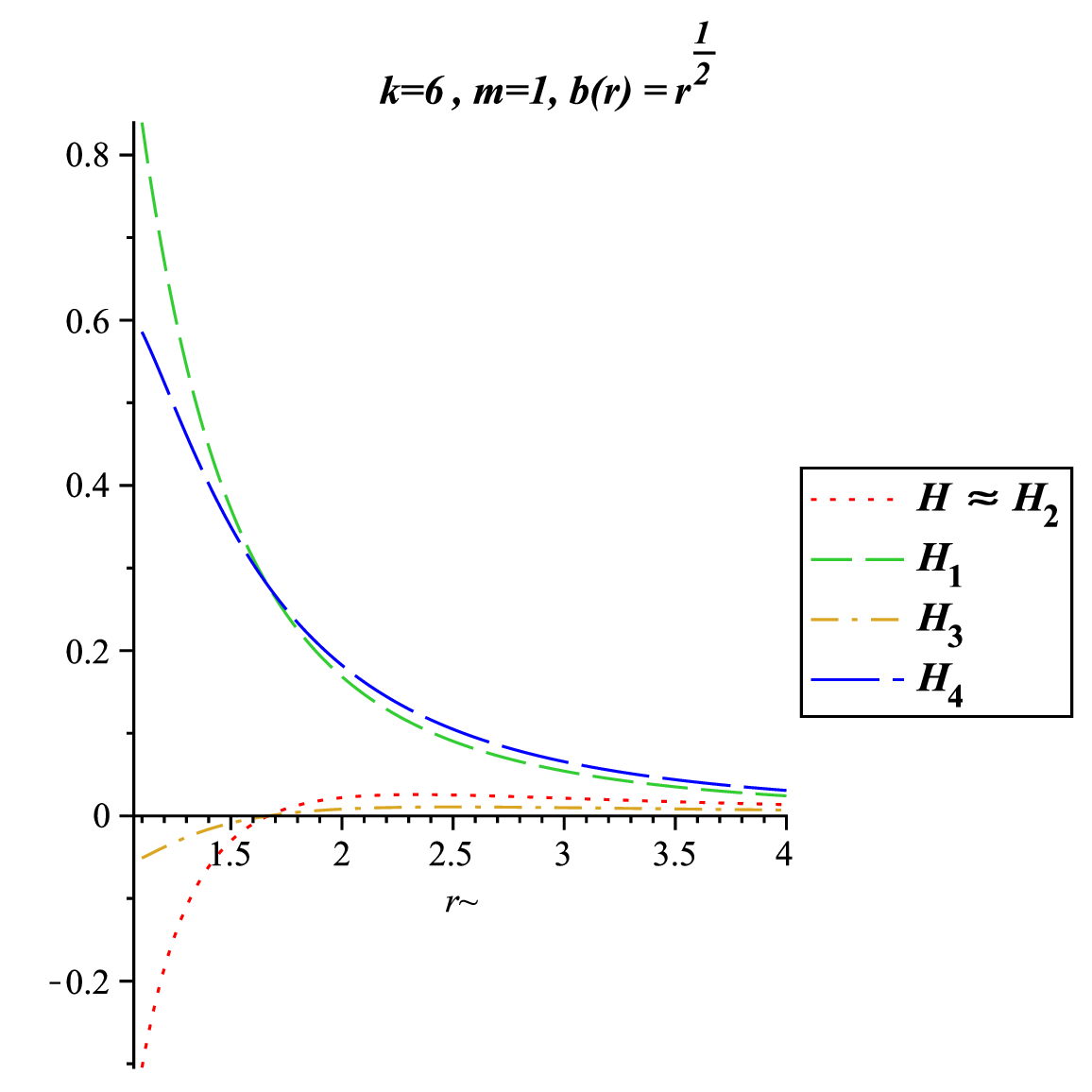}
\caption{The graph displays the functions $H(r)\approx H_2(r)$ (dotted), $H_1(r)$ (dashed), $H_3(r)$ (dash-dotted) and $H_4(r)$ (long-dashed) as functions of the radial coordinate, for the shape function $b(r)=\sqrt{r}$ with $k=6$ and $m=1$. The functions $H$, $H_2$ and $H_3$ are negative only in the vicinity of the throat, whereas $H_1$ and $H_4$ remain positive everywhere. Further details are given in the text.
}
 \label{fig2}
\end{figure}

As an special case we set $\omega_{\infty}=1/2$ to isolate and investigate the effects of the parameter $k$. Substituting the shape function (\ref{11}) alongside functions (\ref{12}) and (\ref{shape23}) for $m = 1$ yields
\begin{equation}
\Phi(r)_6 = -\frac{5}{4\sqrt{r}} - \frac{5}{8r}
\end{equation}
\begin{equation}
\Phi(r)_7 = -\frac{5}{4\sqrt{r}} - \frac{5}{8r} - \frac{1}{3r^{3/2}} - \frac{1}{4r^2}
\end{equation}
\begin{equation}
\Phi(r)_8 = -\frac{5}{4\sqrt{r}} - \frac{5}{8r} - \frac{1}{3r^{3/2}} - \frac{1}{4r^2} - \frac{1}{5r^{5/2}} - \frac{1}{6r^3}
\end{equation}
Note that the indices $6, 7,$ and $8$ correspond to the values of $k$ for the respective cases. Figure~\ref{fig1} illustrates the behavior of  $p_r(r)$, and $p_t(r)$ as functions of the radial coordinate for $b(r)=\sqrt{r}$ and $\Phi(r)_6$. The plots show that  the transverse pressure remain positive throughout the entire spacetime. Meanwhile, the radial pressure $p_r(r)$ is negative near the wormhole throat, but changes sign at a certain radius $r_p > r_0$ in the vicinity of the throat. To examine the ECs for this geometry, we plot the functions $H(r)$, $H_1(r)$, $H_2(r)$, $H_3(r)$, and $H_4(r)$ against $r$ in Fig. \ref{fig2}. As shown in the figure, $H(r)$, $H_2(r)$, and $H_3(r)$ are negative near the wormhole throat, whereas $H_1(r)$ and $H_4(r)$ remain positive everywhere. Crucially, this demonstrates that exotic matter is strictly localized near the throat. Consequently, we conclude that solutions derived from the SV-MCG equation of state significantly reduce the requirement for exotic matter within the framework of GR. This procedure can be similarly applied to other values of $\Phi(r)_k$. While the overall results are identical for all parameters $k$, the radial coordinates at which the signs of $p_r$, $H(r)$, $H_2(r)$, and $H_3(r)$ change vary. These findings are summarized in Table \ref{Tab1}. As shown in the table, the region where exotic matter is confined decreases as $k$ increases.
Let us investigate the SV-MCG equation of state in more details, Equation~(\ref{9f}) relates two unknown functions $\{\rho,p_r\}$ through two free profiles
$\omega_1(r)$ and $\omega_2(r)$.
The term $-\omega_2(r)/\rho$ diverges as $\rho\to0$. This is a singularity of the EoS, not of the metric, but it must be controlled
to keep $p_r(r)$ finite. A weaker alternative is $\omega_2(\infty)=0$, which removes the divergence
at the cost of losing the pure Chaplygin behaviour at the throat. A closer look at the throat and asymptotic limits
Using Eqs.~(\ref{14a}) and (\ref{14b}) in Eq.~(\ref{9f}),
\begin{equation}\label{22}
p_r(r_0)=-\frac{\alpha}{\rho(r_0)},\qquad
p_r(r)\xrightarrow{\;r\to\infty\;}\omega_\infty\,\rho(r).
\end{equation}
Thus the fluid reduces to a pure CG at the throat and to a linear barotropic fluid at spatial infinity (dust for $\omega_\infty=0$,
radiation for $\omega_\infty=1/3$, quasi-vacuum for $\omega_\infty=-1$). This two-regime interpolation is the main conceptual merit of the ansatz.
The flare-out condition of a traversable wormhole requires $\rho+p_r<0$ at $r=r_0$.
With $\omega_1(r_0)=0$, Eq.~(\ref{22}) yields
\begin{equation} \label{23}
\left.\big(\rho+p_r\big)\right|_{r=r_0}=\rho(r_0)-\frac{\alpha}{\rho(r_0)}<0
\;\Longleftrightarrow\;
\alpha>0 \ \text{ and } \ \rho(r_0)<\sqrt{\alpha}.
\end{equation}
Relation~(\ref{23}) provides a quantitative existence window for the wormhole
throat in terms of the SV-MCG parameters, which is absent in the standard (constant)
MCG treatment. Notably, condition (\ref{23}) holds within the range specified in (\ref{11a}).

Let us investigate the behavior of $p_r(r)$, which is composed of two distinct functions
\begin{equation} \label{24}
p_r(r)=p_1(r)+p_2(r)
\end{equation}
 where
 \begin{equation} \label{25}
p_1(r)=\omega_1(r) \rho(r)=(1-\frac{1}{r^m})\omega_\infty\alpha r^{\alpha-3}
\end{equation}
  \begin{equation} \label{26}
p_2(r)=\frac{\omega_2(r)}{ \rho(r)}=\frac{ r^{\alpha-3}}{r^k}.
\end{equation}
 The first function vanishes at the wormhole throat and approaches a linear EoS at extremely large distances. The second function behaves like a pure CG-EoS near the throat and decays to zero at infinity. The combination of these two functions determines the overall behavior of the radial pressure. Near the throat, $p_2$ is the dominant term; however, as $r$ increases, the contribution of $p_1$ grows while that of $p_2$ diminishes. At a specific radial coordinate, $r_p$, these two terms cancel each other out, causing the pressure to change sign. For $r > r_p$, $p_1$ becomes dominant, ensuring that the radial pressure remains positive throughout the outer region. This analysis demonstrates that if $p_2(r)$ approaches zero more rapidly, $r_p$ decreases, thereby restricting the violation of the NEC closer to the throat. Furthermore, relation (\ref{26}) indicates that the rate at which $p_2(r)$ drops to zero increases with $k$. Consequently, one can conclude that larger values of $k$ confine the exotic matter more tightly to the vicinity of the throat. These findings are confirmed by the results presented in Table \ref{Tab1}. A closer inspection of Table \ref{Tab1} reveals that $H_1(r)$ and $H_4(r)$ remain positive across the entire range of the radial coordinate. In contrast, $H(r)$ and $H_2(r)$ are negative near the wormhole throat and change sign at $r_H$. Furthermore, the table indicates that $r_H$ decreases as $k$ increases, implying that the violation of the NEC, WEC, and SEC explicitly depends on $r_H$, whereas the violation of the DEC depends on $r_{H_3}$. Finally, Table \ref{Tab1} shows that the variation of $r_{H_3}$ with respect to $k$ does not follow a regular or predictable pattern.



\section{Concluding remarks}\label{sec4}

In cosmology, the CG and MCG models have drawn significant interest for offering a unified phenomenological framework for dark matter and dark energy, as well as accounting for the universe's late-time accelerated expansion. While the standard CG model transitions from early dust-like behavior to a late-time cosmological constant regime, the MCG model incorporates a linear energy density term to generalize this framework, thereby enhancing flexibility in modeling cosmic fluid evolution. Building on this versatility, we propose a radially dependent generalization of the MCG-EoS to support our wormhole geometry. Unlike conventional CG and MCG models with constant EoS parameters, our formulation allows the pressure-density relationship to vary across the wormhole spacetime. This spatial variation accommodates the inhomogeneous and anisotropic matter distribution inherent to wormholes, as the radially dependent parameters directly influence matter behavior near the throat and in asymptotic regions. Consequently, this generalized MCG-inspired framework serves as a robust phenomenological tool for analyzing the physical properties of the wormhole solution, specifically regarding the satisfaction or violation of ECs.

In this work, we investigated the gravitational field and matter content of a class of wormhole solutions within GR. Our analysis reveals a critical interplay between the geometric parameters and the conditions required for a traversable throat. While selecting an appropriate EoS is essential in classical GR, the fluid supporting the spacetime is often modeled phenomenologically in existing literature, limiting both its physical foundation and flexibility.

We have introduced a generalized, radially dependent EoS of the form $p(r) = \omega_1(r)\rho(r) - \omega_2(r)\rho(r)^{-\gamma}$. Our results reveal that the choice of this EoS is not merely a mathematical convenience but a fundamental necessity for ensuring the consistency of the wormhole solution. We have shown that the pressure at the throat is uniquely determined by the parameters $B$ and $\gamma$, and that a specific consistency relation must be satisfied to align the EoS with the Einstein field equations.
A significant finding of this study is the localization of exotic matter properties. We have demonstrated that the violation of the NEC, a prerequisite for wormhole stability, is primarily governed by the geometric exponent $\alpha$. Specifically, the condition $0<\alpha < 1$ ensures that the NEC is violated at the throat ($r = r_0$), effectively confining the necessary exotic matter behavior to the immediate vicinity of the throat. This localization provides a physically appealing scenario where the non-trivial gravitational effects are concentrated in the high-curvature region, potentially diminishing their impact on the asymptotic structure of spacetime.

Moreover, the asymptotic behavior of the matter content is strictly controlled by the radial functions $\omega_1(r)$ and $\omega_2(r)$. By imposing $\lim_{r \to \infty} \omega_1(r) = \omega_\infty$ and an appropriate decay rate for $\omega_2(r)$, we have successfully constructed a model that transitions from an exotic, non-linear regime at the throat to a well-behaved, barotropic fluid in the far-field limit. This transition ensures that the wormhole remains asymptotically consistent with standard gravitational expectations. Although $\omega_1(r)$ and $\omega_2(r)$ admit a wide range of possible functional forms, we have focused on the specific cases given by (\ref{14a}) and (\ref{14b}). Our analysis demonstrates that the confinement of exotic matter depends explicitly on the choice of free parameters within these functions. Specifically, we showed that as $k$ increases, the exotic matter becomes more tightly confined to the vicinity of the throat. Future research could explore alternative choices for $\omega_1(r)$ and $\omega_2(r)$ to investigate new physical outcomes.

Ultimately, combining the slope of the shape function with a specialized EoS offers a reliable way to construct traversable wormholes. This approach allows for realistic throat modeling while effectively managing the distribution of exotic matter.

\end{document}